# Hybrid-illumination multiplexed Fourier ptychographic microscopy with robust aberration correction

Shi Zhao*, Haowen Zhou, and Changhuei Yang

Department of Electrical Engineering, California Institute of Technology, Pasadena, California 91125, USA

*szhao5@caltech.edu

**Abstract:** Fourier ptychographic microscopy (FPM) is a powerful computational imaging modality that achieves high space–bandwidth product imaging for biomedical samples. However, its adoption is limited by slow data acquisition due to the need for sequential measurements. Multiplexed FPM strategies have been proposed to accelerate imaging by activating multiple LEDs simultaneously, but they typically require careful parameter tuning, and their lack of effective aberration correction makes them prone to image degradation. To address these limitations, we introduce hybrid-illumination multiplexed Fourier ptychographic microscopy (HMFPM), which integrates analytic aberration extraction capability with the efficiency of multiplexed illumination. Specifically, HMFPM employs a hybrid illumination strategy and a customized reconstruction algorithm with analytic and optimization methods. This hybrid strategy substantially reduces the number of required measurements while ensuring robust aberration correction and stable convergence. We demonstrate that HMFPM achieves 1.08 μm resolution, representing a 4-fold enhancement over the system's coherent diffraction limit, across a 1.77×1.77 mm$^2$ field of view using 20 measurements. HMFPM remains robust under diverse aberrations, providing up to 84 μm digital refocusing capability, and effectively corrects both field-dependent and scanning-induced aberrations in whole-slide pathology imaging. These results establish HMFPM as a practical, high-throughput, and aberration-free solution for biological and biomedical imaging.

## 1. Introduction

Fourier ptychographic microscopy (FPM) has emerged as a powerful computational imaging technique for overcoming the space–bandwidth product (SBP) limitation of conventional microscopy [1–5]. By sequentially acquiring intensity images under diverse illumination angles from light-emitting diodes (LEDs) and coherently synthesizing their information in Fourier space, FPM enables quantitative amplitude and phase imaging with high resolution and large field of view simultaneously. These advantages make FPM highly promising for a broad range of biological and biomedical applications [4,6–11]. However, a key limitation of FPM is its relatively slow acquisition speed [9,12–14]. FPM demands high spectral data redundancy to ensure reliable algorithm convergence of the nonconvex optimization problem [15,16], requiring tens to hundreds of low-resolution measurements under different illumination angles. This acquisition overhead restricts the practicality of FPM in dynamic or high-throughput imaging scenarios.

A straightforward and effective strategy to mitigate this limitation is to illuminate multiple LEDs simultaneously within a single capture, thereby reducing the total number of required measurements. One class of these multiplexing approaches exploits the intrinsic degrees of freedom of light and implements multiplexing via hardware modifications to the FPM system. Wavelength-multiplexing FPM illuminates the sample simultaneously with three angle-diverse LEDs of different wavelengths (red, green, and blue) and employs a color camera to decouple one mixed measurement into three monochromatic images [17–20]. However, this technique relies heavily on the assumption of uniform sample response across wavelengths and is prone to chromatic aberrations. It also suffers from anisotropic lateral resolution and has not yet been demonstrated to be effective for dark-field measurements [21]. A related method, polarization-multiplexing FPM, instead encodes distinct illumination angles using different polarization states and decodes them with a polarization-sensitive camera [21–23]. This strategy is constrained by reduced light throughput, stringent requirements for polarization calibration, and increased system complexity, and is typically applicable only to samples without polarization-dependent responses.

Another class of multiplexing approaches overcomes the limitations of wavelength and polarization multiplexing by illuminating the sample with multiple LEDs of identical wavelength and polarization state and decoupling the mixed signals algorithmically without hardware modifications. This class can be broadly divided into two categories: traditional and learning based. Traditional approaches use random or manually designed multiplexed illumination patterns together with customized reconstruction algorithms [12,21,24–28]. For instance, Tian et al. [12] proposed a multiplexed framework in which four LEDs were randomly activated in one measurement and designed an FPM reconstruction algorithm leveraging quasi-Newton second-order optimization to achieve efficient decoupling and image recovery. A subsequent study [24] further employed source-coded illumination, integrating differential phase contrast (DPC) with multiplexed FPM to reduce the number of required measurements. However, these methods often suffer from the nonconvex nature of the reconstruction problem in the multiplexing case, necessitating careful parameter tuning and high-quality initialization to achieve satisfactory reconstruction fidelity. Learning-based approaches train neural networks to optimize multi-LED illumination patterns, thereby enabling efficient sampling and reducing the number of measurements [29–32]. However these methods often highly rely on the training sets and have poor generalization capabilities for different systems and sample types. Moreover, a common limitation of all the aforementioned multiplexing approaches is their lack of, or only very limited, capacity for aberration extraction due to the inherent nonconvexity of the reconstruction problems. As a result, their reconstruction quality is highly susceptible to system aberrations such as defocus, limiting their applications in practical imaging scenarios. To date, none of the reported multiplexing approaches have demonstrated the ability to achieve even moderate aberration correction.

Angular ptychographic imaging with a closed-form method (APIC) is a recently developed analytic framework for high SBP, aberration-free complex-field imaging [33–35]. It initially records measurements under illumination angles matching the objective's maximum acceptance angle (numerical aperture matching (NA-matching) measurements), enabling spectrum reconstruction and pupil function extraction through Kramers–Kronig (K-K) relations [36–39] and analytic aberration extraction algorithms. It then extends higher spatial frequency content by solving linear equations related to the cross-correlation between the known and unknown spectral components using dark-field measurements. Although it shares a similar imaging configuration with conventional FPM, APIC demonstrates superior robustness across varying aberration levels and provides a closed-form complex field solution that eliminates the need for optimization parameter tuning. However, APIC still suffers from long acquisition time, as a large number of dark-field measurements are required to ensure sufficient spectral overlapping [33]. The analytic nature of its reconstruction algorithms further precludes the use of multiplexing strategy to alleviate this limitation, hindering its applicability in practice.

To resolve the problems related to multiplexed FPM and APIC, in this study, we introduce the hybrid-illumination multiplexed Fourier ptychographic microscopy (HMFPM) - a complex-field imaging framework that combines the strengths of multiplexed FPM and APIC. HMFPM operates in two stages. First, eight NA-matching measurements are acquired and processed using K-K relations and analytic aberration extraction method to reconstruct the sample spectrum corresponding to the NA-matching measurements (termed as bright-field spectrum) and estimate the pupil function. Second, a few dark-field measurements are recorded using multiplexed illumination patterns, in which 3–6 LEDs are simultaneously activated with specially designed illumination patterns, the required number of measurements is determined by the severity of system aberrations. Dark-field reconstruction is performed using a customized optimization algorithm, initialized with and constrained by the bright-field spectrum and the previously extracted pupil function. HMFPM significantly reduces the number of required measurements while retaining robust aberration extraction capability. Since the pupil function is analytically determined and the reconstruction is initialized with a high-quality, aberration-corrected bright-field spectrum, the customized optimization algorithm converges more rapidly and stably, without the need for tuning relaxation factors. Simulations and experiments validated that, with a 4× objective, HMFPM attains 1.08 μm resolution, representing a 4-fold improvement over the system's coherent diffraction limit, across a 1.77×1.77 mm$^2$ field of view (FOV). This performance remains robust under diverse aberrations including up to 84 μm defocus aberration in experiments, while requiring fewer or a comparable number of measurements than previous angle-multiplexing approaches [12,24]. Furthermore, we demonstrate that HMFPM effectively corrects field-dependent and scanning-induced aberrations in whole-slide pathological imaging, achieving image quality comparable to APIC but with substantially fewer measurements. We believe HMFPM provides a more efficient, robust, and aberration-free solution for high-throughput, high-resolution and wide-field imaging.

## 2. Methods

*2.1 Imaging setup and forward propagation model of HMFPM*

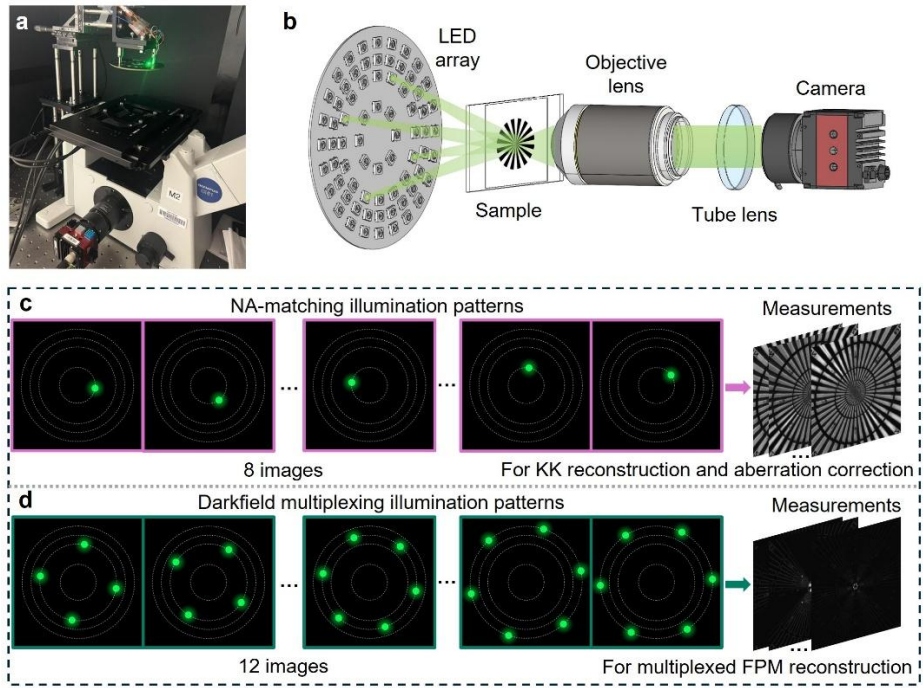

Figure 1. Experimental setup and illumination strategy for HMFPM. (a) A photo of the imaging system setup. (b) Schematic diagram of the optical configuration. (c) Illumination patterns for NA-matching measurements. (d) Illumination patterns for dark-field measurements

HMFPM employs an optical configuration similar to that of conventional FPM. As illustrated in Fig. 1(a)(b), a programmable LED disk is positioned in front of the sample to provide quasi-plane-wave illumination from various oblique angles. The sample is then imaged using a standard 4f microscopy system including an objective lens, a tube lens, and a camera.

During the data acquisition, for each measurement, one or multiple LEDs are switched on to illuminate the sample, and the camera records a corresponding low-resolution image. The forward process from illumination to image acquisition can be modeled as follows: the illumination wavefront from the $m^{th}$ LED ($m = 1, 2, ..., N$) is approximated by a tilted plane wave

$$u(\mathbf{r}; \mathbf{k}_m) = e^{j\mathbf{k}_m \cdot \mathbf{r}}, \tag{1}$$

where $\mathbf{k}_m$ is the transverse illumination wavevector and $\mathbf{r}$ represents the lateral spatial coordinates. This plane wave interacts with sample $o(\mathbf{r})$, and the resulting optical field is transmitted through the 4f system. The modulated sample spectrum (also known as auxiliary function) $\hat{S}_m$ at the camera plane can be written as

$$\hat{S}_m(\mathbf{k}) = \hat{O}(\mathbf{k} - \mathbf{k}_m)\hat{H}(\mathbf{k}) = \hat{O}(\mathbf{k} - \mathbf{k}_m)\text{Circ}_{\text{NA}}(\mathbf{k})e^{j\hat{\phi}(\mathbf{k})}, \tag{2}$$

where $\hat{O} = \mathcal{F}\{o(\mathbf{r})\}$ represents the sample's spectrum, $\mathcal{F}$ is the two-dimensional (2D) Fourier transform operator, $\mathbf{k}$ is the spatial frequency coordinate vector, and $H$ is the pupil function (also referred to as the coherent transfer function in coherent imaging) of the imaging system. The pupil function is described by a circular function $\text{Circ}_{\text{NA}}$ with an NA-dependent radius, combined with an aberration phase term $\phi$. With the objective NA specified, the pupil function can be completely characterized by the aberration phase term $\phi$. Due to the Fourier transform property, varying the illumination angle laterally shifts the sample spectrum in Fourier space, enabling sampling of different portions of it. The camera records only the intensity of the complex-valued optical field, expressed as

$$I_m(\mathbf{r}) = \left|\mathcal{F}^{-1}\{\hat{S}_m\}\right|^2 = \left|\left[\mathcal{F}^{-1}\{\hat{O}(\mathbf{k} - \mathbf{k}_m)\text{Circ}_{\text{NA}}(\mathbf{k})e^{j\hat{\phi}(\mathbf{k})}\}\right]\right|^2, \tag{3}$$

where $\mathcal{F}^{-1}$ denotes the inverse 2D Fourier transform operator.

When multiple LEDs are illuminated at the same time, the resulting image is formed as the incoherent sum of the intensity contributions from each active LED,

$$I_n(\mathbf{r}) = \sum_{m \in \mathcal{M}_n} \left|\left[\mathcal{F}^{-1}\{\hat{O}(\mathbf{k} - \mathbf{k}_m)\text{Circ}_{\text{NA}}(\mathbf{k})e^{j\hat{\phi}(\mathbf{k})}\}\right]\right|^2, \tag{4}$$

where $\mathcal{M}_n$ represents the set of LEDs used in the $n$-th multiplexed illumination.

## 2.2 Hybrid illumination strategy of HMFPM

HMFPM employs a hybrid illumination strategy that differentiates between bright-field (illumination angles no larger than the objective's acceptance angle) and dark-field (illumination angles larger than the objective's acceptance angle) LEDs. For bright-field illumination, as shown in Fig. 1(c), eight LEDs positioned at angles exactly matching the objective's maximum acceptance angle are sequentially activated to acquire eight NA-matching measurements. Such illumination has been demonstrated to retain the maximum amount of information, particularly for low-frequency phase components, and serves as a prerequisite for applying the K-K relations [38] and analytic aberration extraction [33]. The selection of eight measurements represents a balance between achieving accurate aberration extraction and maintaining reasonable acquisition speed.

For dark-field imaging, customized multiplexed illumination patterns are implemented to efficiently sample the high-frequency regions of the object spectrum as shown in Fig. 1(d). The dark-field LEDs are arranged in three concentric rings (the white dashed rings in Fig. 1(d)), extending the maximum illumination NA up to three times that of the objective. Within each ring, LEDs were divided into groups based on a predefined multiplexing number, with each group comprising LEDs at equal angular intervals (e.g., every $n$-th LED). The groups were then alternately activated, so that in each exposure one evenly spaced set of LEDs was illuminated simultaneously. The acquisition sequence proceeds systematically from the innermost to the outermost ring, thereby yielding comprehensive dark-field spectral coverage.

The dark-field multiplexed patterns are designed with the following considerations. First, annular illumination patterns are employed to ensure isotropic resolution [9]. Second, the multiplexed LED groups are illuminated in an outward radial sequence, with the illumination angle increasing progressively. Compared with random multiplexed illumination patterns, our patterns ensure a gradual expansion of the sampled spectrum, with illumination energy shifting smoothly from low to high spatial frequencies, thereby promoting more stable reconstruction and faster convergence [40]. Third, we alternately illuminated LED groups formed by LEDs positioned at equal angular intervals, which reduces overlap between the sampled spectra of multiplexed LEDs in one measurement. The spectral overlap within one measurement has been demonstrated to degrade reconstruction quality [27]. Meanwhile, this design also broadens the spectral coverage of each measurement and enhances spectral redundancy across different measurements.

## 2.3 Analytic spectrum and pupil reconstruction using NA-matching measurements

A general pipeline of HMFPM reconstruction is illustrated in Fig. 2. The workflow comprises two main stages: (i) Analytic spectrum and pupil reconstruction with eight NA-matching measurements, and (ii) multiplexed reconstruction for spectrum extension with dark-field measurements. In the first stage, the eight NA-matching images are utilized to recover the bright-field spectrum via the K-K relations, thereby enabling aberration extraction and correction. The corrected spectrum, together with the estimated pupil function, is subsequently employed as the initialization and prior for a specially designed and robust multiplexed reconstruction, which leverages dark-field measurements to extend the reconstructed spectrum.

The first step of processing NA-matching measurements is to apply K-K relations to reconstruct the bright-field spectrum. The K-K relations state that the real and imaginary parts of any complex function that is analytic in the upper half-plane are connected through Cauchy integration, allowing one component to be fully determined from the other. Recently, K-K relations have been adopted and shown to be highly effective in complex-field imaging [36–38]. To apply them in our system, the illumination angles must be precisely matched to the objective's maximum acceptance angle (NA-matching) [33,38]. Under this condition, we can define a complex function with a known real part, from which the aberrated sample spectra can be retrieved through Hilbert transform.

For the $m^{th}$ NA-matching measurement ($m = 1, 2, ..., 8$) with illumination wavevector $\mathbf{k}_m$, we first define a spectrally shifted modulated sample spectrum,

$$S'_m(\mathbf{r}) = \mathcal{F}^{-1}\{\hat{S}_m(\mathbf{k} + \mathbf{k}_m)\}, \tag{5}$$

where $\hat{S}_m(\mathbf{k})$ is the modulated sample spectrum defined by Eq. (2). According to the property of Fourier transform,

$$|S'_m(\mathbf{r})| = |\mathcal{F}^{-1}\{\hat{S}_m(\mathbf{k} + \mathbf{k}_m)\}| = |S_m(\mathbf{r})e^{-2\pi \mathbf{k}_m \cdot \mathbf{r}}| = |S_m(\mathbf{r})| = \sqrt{I_m(\mathbf{r})}. \tag{6}$$

We then define a supporting function by taking the logarithm of the modulated sample spectrum and adding a constant phase term $e^{-j\phi(\mathbf{k}_m)}$ to correct the phase shift for the unscattered light, yields

$$A_m(\mathbf{r}) = \log[S'_m(\mathbf{r})e^{-j\theta_m}] = \frac{1}{2}\log[I_m(\mathbf{r})] + j\{S'_m(\mathbf{r}) - \theta_m\}, \tag{5}$$

where $\theta_m = \hat{\phi}(\mathbf{k}_m)$ is a constant phase offset defined by the pupil phase at $\mathbf{k} = \mathbf{k}_m$. It can be proved that $A_m(\mathbf{r})$ satisfies the prerequisite of K-K relations and the Hilbert transform can be used instead of direct Cauchy integration to calculate full $A_m(\mathbf{r})$ [Cite],

$$\hat{G}_m(\mathbf{k}) := \mathcal{F}\{A_m(\mathbf{r})\}(\mathbf{k}) = \begin{cases} [\mathcal{F}(\log I_m)](\mathbf{k}), & \mathbf{k} \cdot \mathbf{k}_m < 0 \\ \frac{1}{2}[\mathcal{F}(\log I_m)](\mathbf{k}), & \mathbf{k} \cdot \mathbf{k}_m = 0 \\ 0, & \mathbf{k} \cdot \mathbf{k}_m > 0 \end{cases} \tag{6}$$

Then, the modulated spectrum can be restored using the inverse Fourier transform and exponential function.

$$\hat{S}'_m(\mathbf{k})e^{-j\theta_m} = \mathcal{F}\{S'_m(\mathbf{r})e^{-j\theta_m}\} = \mathcal{F}\{\exp[\mathcal{F}^{-1}(G_m(\mathbf{k}))]\}. \quad (7)$$

The left-hand side in Eq. (7) includes aberration terms. To handle these aberrations, we worked in the spatial frequency domain and utilized the phases of multiple reconstructed spectra. Eq. (7) can be rewritten as,

$$\hat{S}'_m(\mathbf{k})e^{-j\theta_m} = \hat{A}(\mathbf{k})e^{j\hat{\alpha}(\mathbf{k})}\text{Circ}_{NA}(\mathbf{k}+\mathbf{k}_m)e^{j\hat{\phi}(\mathbf{k}+\mathbf{k}_m)-j\theta_m}, \quad (8)$$

where $\hat{A}(\mathbf{k})$ represents the amplitude of the sample's spectrum and $\hat{\alpha}(\mathbf{k})$ is the phase term. For two different LED illumination angles, we define

$$\mathcal{S}_{ml} \coloneqq \{k|\text{Circ}_{NA}(\mathbf{k}+\mathbf{k}_m)\text{Circ}_{NA}(\mathbf{k}+\mathbf{k}_l) \neq 0\}. \quad (9)$$

$\mathcal{S}_{ml} \neq \emptyset$ when the two parts of the spectra have overlap regions. Within the overlapped region $\mathcal{S}_{ml}$, the phase difference between the two spectra in Eq. (8) yields:

$$\arg(\hat{S}'_m(\mathbf{k})e^{-j\theta_m}) - \arg(\hat{S}'_l(\mathbf{k})e^{-j\theta_l}) = [\phi(\mathbf{k}+\mathbf{k}_m) - \phi(\mathbf{k}+\mathbf{k}_l)] - [\hat{\phi}(\mathbf{k}_m) - \hat{\phi}(\mathbf{k}_l)]. \quad (10)$$

Eq. (10) shows that the contribution from the sample spectrum canceled out, and the difference depended solely on the aberration of the system. The remaining phase difference was linear with respect to the aberration functions. By analyzing all overlapping regions (suppose $M$ in total), we can formulate a linear equation and solve the system aberration term accordingly.

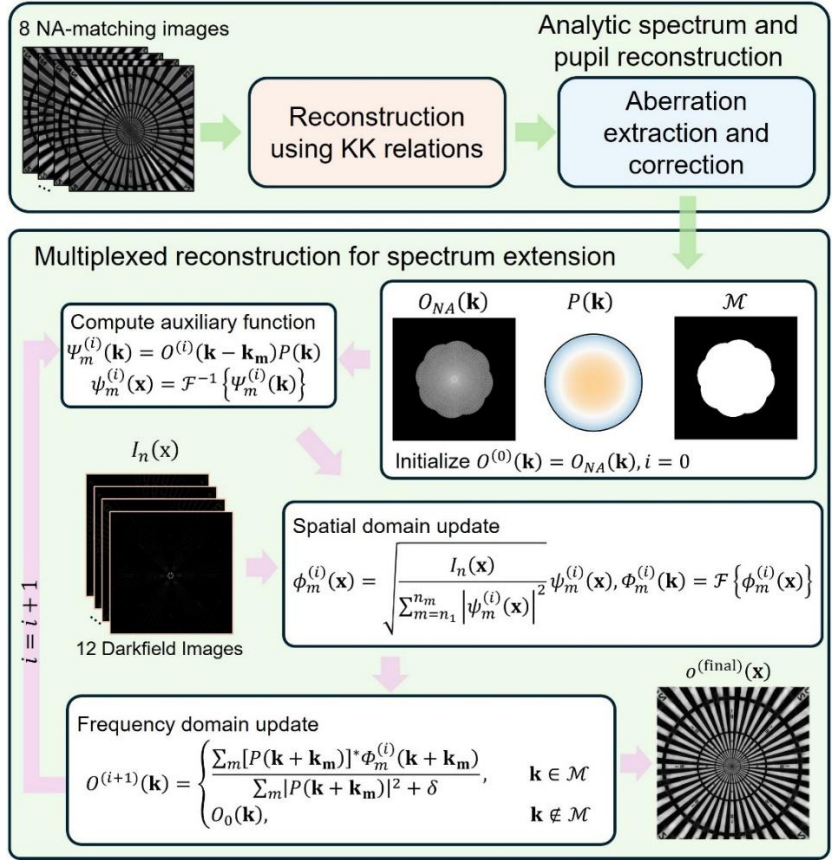

Figure 2. Flow chart of the HMFPM reconstruction pipeline. Eight NA-matching measurements are first used to recover the object spectrum via K-K relations, enabling aberration estimation and correction. The corrected spectrum and the estimated pupil function then serve as initialization for robust multiplexed reconstruction for spectrum extension using 12 dark-field images。

The extracted aberration is applied to correct the modulated spectrum in Eq. (7), yielding an aberration-free reconstruction of the sample bright-field spectrum through spectrum stitching. The pupil function can also be fully determined from the aberration phase term. The corrected spectrum, together with the pupil function, is subsequently used as the initialization and prior for the multiplexed reconstruction for spectrum extension.

*2.4 Multiplexed reconstruction for spectrum extension using dark-field measurements*

A persistent challenge in FPM reconstruction is its inherently nonconvex nature, which makes optimization susceptible to local minima convergence, particularly under multiplexed illumination conditions [2]. Using proper initialization has been proven to be an effective strategy to alleviate this issue and has consequently become an important subject of investigation in FPM research [41,42]. A widely adopted practice is to use a low-resolution intensity image, typically obtained under normal illumination, as the initial estimate [1,4,5,43]. However, such intensity-based initialization fails to preserve low-frequency phase information, often resulting in degraded reconstructions [44,45]. Tian *et al.* [24] proposed using the linearly approximated phase solution from DPC deconvolution as a close initial guess, but the DPC spectrum itself still suffers from poor low-frequency response and anisotropic resolution. These limitations underscore the necessity for more reliable and robust initialization strategies in multiplexed FPM reconstruction.

Our proposed framework, HMFPM, provides a stable and effective solution to this challenge. Specifically, HMFPM applies analytic methods to NA-matching measurements, yielding a high-quality, aberration-free bright-field spectrum. This spectrum provides an ideal initialization for subsequent multiplexed dark-field reconstruction, mitigating the limitations of previously reported initialization strategies. Furthermore, the aberration extracted from analytic stages provides an accurate estimate of the pupil function, which can be directly incorporated as prior knowledge in the FPM reconstruction process. This obviates the need for joint sample–pupil optimization, a complex and computationally demanding step in conventional FPM, thereby accelerating convergence while enhancing stability. Building upon this initialization and prior information, we designed a dedicated multiplexed reconstruction algorithm, whose mathematical derivation and implementation details are presented below.

For clarity, we denote $\hat{O}_0(\mathbf{k})$ as the reconstructed bright-field spectrum, and $\hat{H}(\mathbf{k})$ as the extracted pupil function. Since the analytic aberration extraction method has demonstrated superior reconstruction fidelity and robustness [33], both $\hat{O}_0(\mathbf{k})$ and $\hat{H}(\mathbf{k})$ remain fixed throughout the reconstruction process. In addition, we define a spectral mask $\mathcal{S}$, which specifies the nonzero region of $\hat{O}_0(\mathbf{k})$, corresponding to the effective bright-field spectrum coverage.

The reconstruction of high-frequency sample spectrum from multiplexed dark-field measurements can be formulated as a least-squares optimization problem, with an objective function derived from the forward model in Eq. (4),

$$\min_{\hat{O}(\mathbf{k})} \sum_{n=1}^{N} \sum_{\mathbf{r}} \left| I_n^{ms}(\mathbf{r}) - \sum_{m \in \mathcal{M}_n} \left|[\mathcal{F}^{-1}\{\hat{O}(\mathbf{k}-\mathbf{k}_m)\hat{H}(\mathbf{k})\}]\right|^2 \right|^2, \text{s.t.} \, \hat{O}(\mathbf{k}) = \hat{O}_0(\mathbf{k}), \mathbf{k} \in \mathcal{S} \quad (11)$$

where $\mathcal{M}_n$ represents the set of LEDs used in the *n*-th multiplexed illumination, $N$ is the total number of dark-field measurements and $I_n^{ms}(\mathbf{r})$ is the *n*-th intensity measurement. Unlike conventional multiplexed FPM reconstruction [12], $\hat{H}(\mathbf{k})$ is no longer required for optimization and $\hat{O}(\mathbf{k})$ is constrained to $\hat{O}_0(\mathbf{k})$ within the spectral support $\mathcal{S}$.

To facilitate optimization of Eq. (11), we adopt the alternating projection strategy, which decomposes the optimization problem into two tractable subproblems: a spatial-domain projection enforcing the measured intensity constraint, and a frequency-domain projection ensuring spectral consistency with the prior spectrum constraint.

Consider the *i*-th iteration for the *n*-th measurement, we introduce an intermediate spatial variable $\phi_m(\mathbf{r})$ to minimize the discrepancy between the measured intensity and the estimated intensity from the (*i*−1)-th iteration,

$$\left\{\phi_m^{(i)}(\mathbf{r})\right\} = \arg\min_{\phi_m(\mathbf{r})} \sum_{m \in \mathcal{M}_n} \left|\phi_m^{(i)}(\mathbf{r}) - \mathcal{F}^{-1}\{\hat{O}^{(i-1)}(\mathbf{k} - \mathbf{k}_m)\hat{H}(\mathbf{k})\}\right|^2,$$
$$\text{s.t.} \ I_n(\mathbf{r}) = \sum_{m \in \mathcal{M}_n} \left|\phi_m^{(i)}(\mathbf{r})\right|^2. \quad (12)$$

This constrained optimization is solved via the Lagrangian multiplier method, yielding the classical intensity-projection update,

$$\phi_m^{(i)}(\mathbf{r}) = \sqrt{\frac{I_n(\mathbf{r})}{\sum_{m \in \mathcal{M}_n}\left|s_m^{(i-1)}(\mathbf{r})\right|^2}}\, s_m^{(i-1)}(\mathbf{r}), \Phi_m^{(i)}(\mathbf{k}) = \mathcal{F}\left\{\phi_m^{(i)}(\mathbf{r})\right\}, \quad m \in \mathcal{M}_n \quad (13)$$

where $s_m^{(i-1)}(\mathbf{r}) = \mathcal{F}^{-1}\{\hat{O}^{(i-1)}(\mathbf{k} - \mathbf{k}_m)\hat{H}(\mathbf{k})\}$, and $\Phi_m^{(i)}(\mathbf{k})$ is the Fourier transform of $\phi_m^{(i)}(\mathbf{r})$.

In the Fourier domain, the sample spectrum is then updated using $\Phi_m^{(i)}(\mathbf{k})$, which incorporates the spatial-domain intensity constraint, while simultaneously enforcing the prior spectral support constraint of $O_0(\mathbf{k})$. This leads to the following optimization,

$$\left\{\hat{O}^{(i)}(\mathbf{k})\right\} = \arg\min_{O(\mathbf{k})} \sum_{m \in \mathcal{M}_n} \left|\hat{O}(\mathbf{k} - \mathbf{k_m})\hat{H}(\mathbf{k}) - \Phi_m^{(i)}(\mathbf{k})\right|^2, \ \text{s.t.} \ \hat{O}(\mathbf{k}) = \hat{O}_0(\mathbf{k}), \mathbf{k} \in \mathcal{S}, \quad (14)$$

where $\hat{H}(\mathbf{k})$ remains fixed as a prior knowledge, and only $\hat{O}(\mathbf{k})$ is updated. The optimization can still be solved via simple Lagrangian multiplier method,

$$O^{(i+1)}(\mathbf{k}) = \begin{cases} \dfrac{\sum_{m \in \mathcal{M}_n}[\hat{H}(\mathbf{k} + \mathbf{k_m})]^* \Phi_m^{(i)}(\mathbf{k} + \mathbf{k_m})}{\sum_{m \in \mathcal{M}_n}|\hat{H}(\mathbf{k} + \mathbf{k_m})|^2 + \beta}, & \mathbf{k} \in \mathcal{N} \\ O_0(\mathbf{k}), & \mathbf{k} \notin \mathcal{N} \end{cases}, \quad (15)$$

where $\beta$ is a fixed small value regularization constant. It can be observed from Eq. (15) that no relaxation factor (step size) is required for spectrum updates. As the relaxation factor is typically the most sensitive and critical parameter in conventional FPM, this eliminates the need for extensive parameter tuning, thereby improving scalability and ensuring reproducibility. Compared with the classic Gauss–Newton method, our approach demonstrates superior stability, enhanced robustness to noise, and reduced computational complexity.

The workflow of our customized multiplexed reconstruction algorithm is as follows. First, $\hat{H}(\mathbf{k})$ is fixed as pupil function prior and $\hat{O}_0(\mathbf{k})$ is used as the initialization $\hat{O}^{(0)}(\mathbf{k})$ of the sample spectrum. During the iterative process, Eq. (13) and Eq. (15) are applied alternately to update the spectrum until convergence is achieved with sufficient accuracy. A schematic flowchart of the proposed algorithm is provided in the bottom panel of Fig. 2.

## 3. Results

### 3.1 Simulation verification of HMFPM

To validate the performance of the proposed HMFPM method, we conducted comprehensive simulations using a pure-phase Siemens star for resolution quantification and natural images for reconstruction quality evaluation. All simulations were performed with a 4×/0.13 NA low-magnification objective lens, a camera pixel size of 3.45 μm, and an LED central wavelength of 516.2 nm. For HMFPM, unless otherwise specified, we followed the illumination strategy described in Section 2.2: eight NA-matching illuminations were applied at incidence angles close to the maximum acceptance angle of the objective, and twelve dark-field illuminations were distributed across three concentric rings at illumination angles of 16°, 20°, and 23°. Each ring contained four LEDs at equal angular intervals, and multiplexing was performed by simultaneously activating 4, 6, and 6 LEDs from the inner, middle, and outer rings, respectively, to extend the accessible spectrum. We benchmarked HMFPM against the state-of-the-art complex imaging technique APIC [33], as well as two widely recognized multiplexing strategies: multiplexed coded illumination FPM (MFPM) [12] and computational illumination FPM (SFPM) [24] proposed by Lei et al. For APIC, sequential illuminations with 88–96 measurements were employed to ensure sufficient spectral overlap. MFPM and SFPM were implemented using the original codes and parameter configurations provided in the respective publications, with the number of measurements chosen to match the spectrum overlap rates reported in those studies. To ensure a fair comparison for resolution quantification, MFPM and SFPM were also configured with annular illumination. In addition, SFPM implemented the updated DPC algorithm with computational aberration correction capacity [46], enabling a more objective evaluation of its aberration-correction capability.

We first quantified the resolution achieved by HMFPM and other methods under varying aberration conditions using a simulated pure-phase Siemens star target. To quantitatively assess the optical resolution, we identified the smallest radius at which all peaks along the circular line profile were preserved, corresponding to the sparrow limit [47]. As shown in Figs. 3(a1–a4), in the absence of aberrations, all four methods reached an optical resolution of 1.02 μm, with the corresponding resolution circles highlighted and the radial intensity profiles along these circles displayed as insets to the right of each panel. Even under this condition, HMFPM excels by requiring fewer measurements, as the spectrum initialization and pupil priors provided by analytic reconstruction mitigate the need for the high spectral overlap typically demanded in conventional multiplexed FPM. Simulations under moderate and large arbitrary aberrations are shown in Figs. 3(b1–b4) and Figs. 3(c1–c4), respectively. With only 20 measurements, HMFPM consistently maintained the same resolution and produced high-quality reconstructions comparable to the aberration-free case and to APIC using 96 measurements. Notably, the aberrations extracted by HMFPM closely matched the ground truth, as displayed in the left panel. In contrast, MFPM and SFPM exhibited degraded performance: under moderate aberrations, the reconstructions became noisy with clear artifacts and reduced resolution (Figs. 3(b3)–3(b4)), and under large aberrations, the reconstructions failed entirely (Figs. 3(c3)–3(c4)).

Figs. 3(d1–d4) further illustrate the performance of the four methods under a severe, 80-μm defocus-dominant aberration. Defocus aberrations impose strong modulation across the entire spectrum, rendering the reconstruction problem ill-conditioned and more difficult to solve compared with other aberrations [48]. In this challenging case, HMFPM maintained the target resolution by slightly reducing the number of multiplexed LEDs in one single measurement and using 28 measurements, whereas MFPM and SFPM failed even when the number of measurements was proportionally increased.

To more comprehensively evaluate the performance of HMFPM, we moreover conducted simulations using natural images as complex-field samples and quantified reconstruction quality with structural similarity index (SSIM) values [49] relative to the ground truth under varying aberration conditions. As shown in Fig. 4(a), when no aberration was present, all four methods produced relatively good reconstructions of both amplitude and phase. Under large random (Fig. 4(b)) and defocus-dominant aberrations (Fig. 4(c)), HMFPM achieved high-quality complex-field reconstructions with 20 measurements, maintaining SSIM values above 0.89 for both amplitude and phase reconstruction. The results are comparable to APIC, but HMFPM used substantially fewer measurements. In contrast, MFPM and SFPM essentially failed under these conditions. This simulation further highlights the robustness of HMFPM to aberrations, demonstrating a clear advantage over other multiplexing methods.

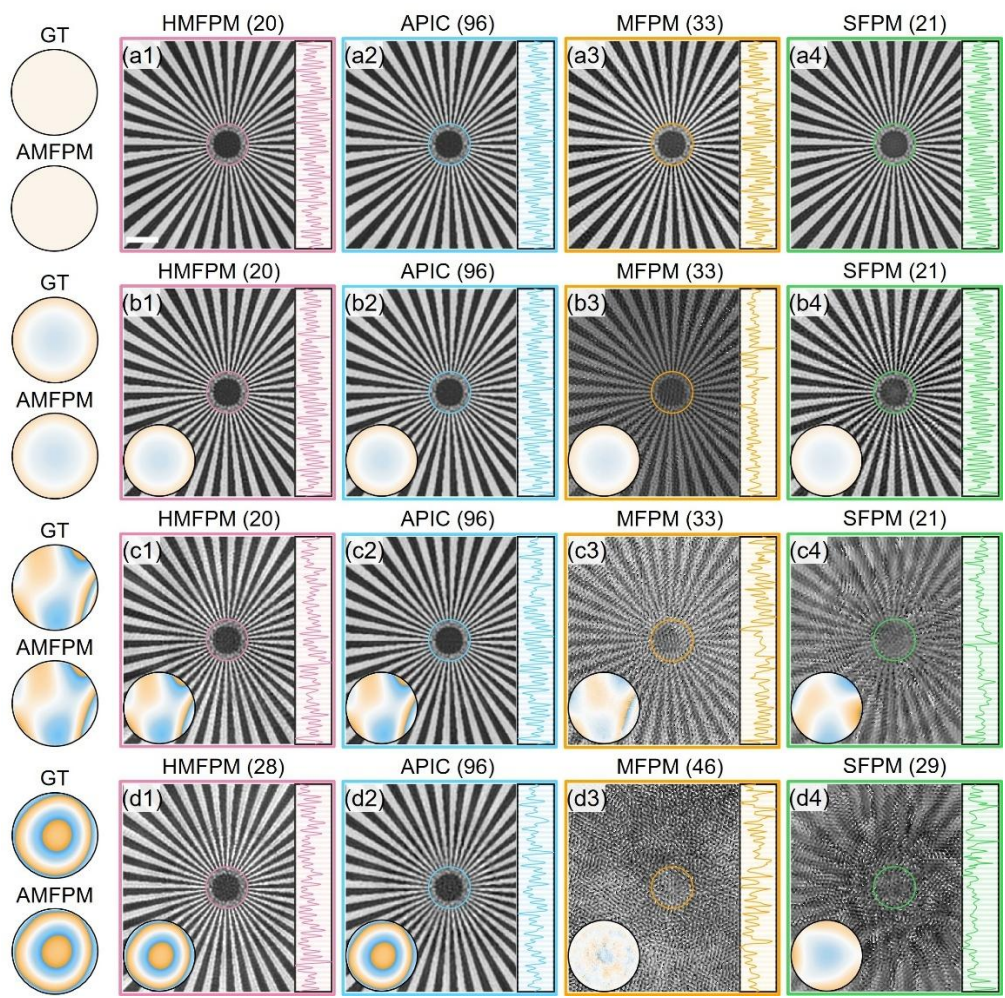

Figure 3. Comparison of different reconstruction methods under varying aberration conditions. Rows (a–d) represent increasing levels and types of aberrations, with the ground truth (GT) aberration phase maps shown on the left. Columns (1–4) display the results of (1) HMFPM (proposed), (2) APIC, (3) MFPM, and (4) SFPM. The numbers in brackets indicate the number of measurements used by each method. Insets highlight the recovered aberrations and the radial intensity profiles along the marked circles (corresponding to the achieved resolution limits). Scale bar 10 μm.

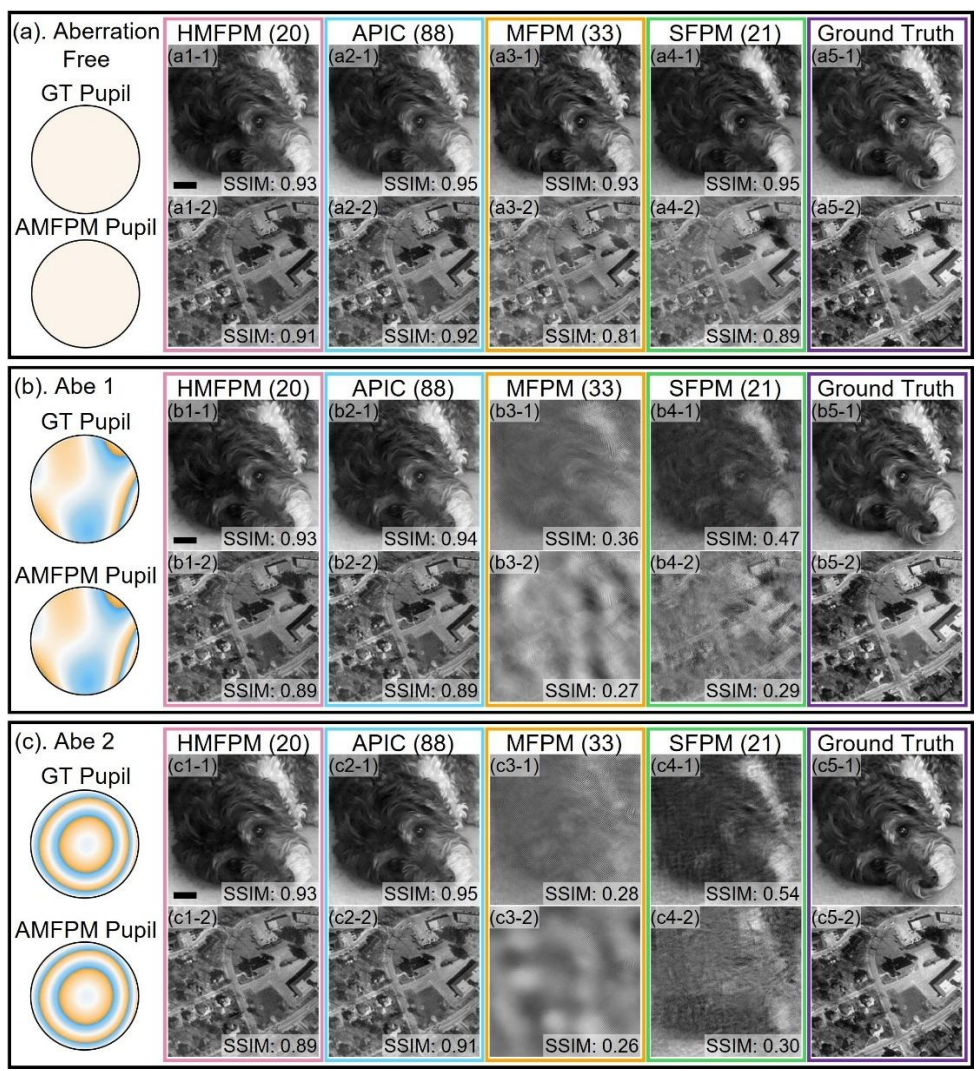

Figure 4. Comparative reconstructions of natural images under different aberration conditions. (a) Aberration-free, (b) hybrid aberration (Abe 1), and (c) defocus-dominated aberration (Abe 2). Columns (1–4) correspond to the reconstructed amplitude(x-1) and phase (x-2) results from (1) HMFPM (20 measurements), (2) APIC (88 measurements), (3) MFPM (33 measurements), and (4) SMFPM (21 measurements). Column (5) shows the ground truth images. Structural similarity index (SSIM) values with respect to ground truth are annotated at the lower right of each image. Scale bar 20 μm.

*3.2 Experiments quantification using a Siemens star*

We then imaged a standard Siemens star target and pathological samples to experimentally validate the performance of HMFPM. All experiments listed below were performed using a 4×/0.13 NA objective lens (Plan N, Olympus) mounted on an Olympus IX51 microscope body and coupled with a camera (Allied Vision Prosilica GT6400). The camera had a pixel pitch of 3.45 μm, with each FOV consisting of 2048 × 2048 pixels, corresponding to a physical size of 1.77 mm × 1.77 mm. Illumination was provided by a customized, programmable LED array (Green: XQAGRN-00-0000-000000Y01, 520 nm wavelength), controlled by an Arduino microprocessor (Arduino Uno). The illumination strategies used for HMFPM and APIC in experiments followed the same configurations as described in Section 3.1 for simulations.

In our first experiment, we used a Siemens star target to quantify the resolution achieved by HMFPM under different levels of aberration. We focused on defocus-dominant aberrations, as defocus is the most prevalent aberration encountered in practical imaging systems. The sample was manually displaced to specific defocus distances up to 84 μm, after which measurements were acquired and HMFPM reconstructions were performed. As the defocus distance increased, the reconstruction problem became increasingly ill-posed. Accordingly, we gradually increased the total number of measurements from 20 to 28 to ensure good results. HMFPM reconstructions at different defocus distances are shown in Figs. 5(a–d), with the corresponding extracted aberration shown as insets. For reference, the in-focus APIC reconstruction using 88 measurements [Fig. 5(e)] and the bright-field microscope images acquired with a 20×/0.4 NA objective lens [Fig. 5(f)] are provided as ground truth. Similar to the simulation, we identified the smallest radius at which 40 peaks along the circular line profile were preserved to quantitatively assess the optical resolution. These radii are highlighted with circles, and the corresponding radial profiles are displayed to the right of each image. We observed that even at defocus distances up to 84 μm, HMFPM successfully resolved the stripes, achieving an optical resolution of 1.08 μm, close to the theoretical resolution of 1.00 μm (corresponding to 0.52 NA). The reconstruction results are comparable to those obtained with in-focus bright-field microscopy and APIC.

In addition to reducing acquisition time, HMFPM also substantially outperforms APIC in reconstruction speed. Figure 3(g) presents a runtime comparison between the HMFPM algorithm and the GPU-accelerated APIC algorithm [35] as a function of image patch size, using the in-focus Siemens star target. Both algorithms were tested on a desktop with an Intel Core i7-14700KF CPU and an NVIDIA GeForce RTX 4090 GPU, with the runtime for each configuration averaged over five trials. As shown in Fig. 3(f), HMFPM consistently required significantly less reconstruction time than APIC across all patch sizes, with the performance gap widening as the patch size increased. Specifically, for a patch size of 512 pixels, APIC required approximately 125 s, whereas HMFPM completed the reconstruction in only 1.73 s, corresponding to a 72-fold speed improvement. Moreover, APIC failed to reconstruct images at patch sizes larger than 512 due to GPU memory overflow, while HMFPM remained computationally efficient, requiring only 5.18 s at a patch size of 1024.

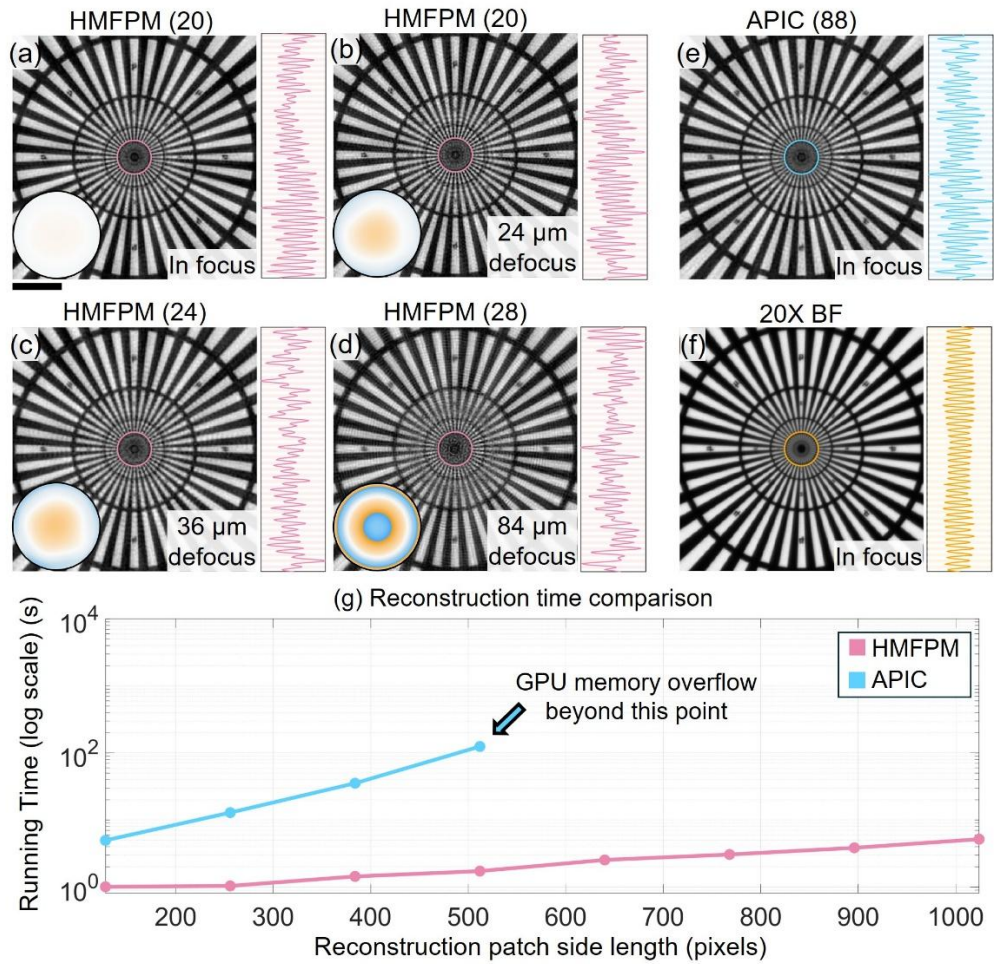

Figure 5. HMFPM Reconstruction results of Siemens star under different aberration levels and computational efficiency comparison between HMFPM and APIC. The numbers in brackets indicate the number of measurements used. (a–d) HMFPM reconstructions under different defocus conditions: (a) in-focus, (b) 24 μm defocus, (c) 36 μm defocus, and (d) 84 μm defocus. (e) In-focus APIC reconstruction. (f) Standard 20× 0.4 NA bright-field (BF) image served as ground truth. (g) Comparison of reconstruction running time (log scale) as a function of patch size. Scale bar 30 μm.

*3.3 Phase imaging of unstained human lung cancer section with HMFPM*

Phase imaging offers a promising approach to mitigate the issue of stain variation in deep learning downstream analysis [50], to help rescue decades-old pathology slides [51], and to provide structural information for pathological assessments [52,53]. Here, we demonstrate the capacity and effectiveness of HMFPM for high-throughput, high-quality pathological phase imaging. An unstained human non-small-cell lung cancer pathology slide was imaged using HMFPM with 20 measurements. Compared with multiplexed FPM approaches using similar amount of measurements, HMFPM excels in effectively correcting system aberrations to achieve high-quality, aberration-free phase imaging. Figure 6(a) shows a high-quality reconstruction of one FOV obtained with HMFPM. Because the system FOV is large, field-dependent aberrations arise, varying across different spatial locations of the field of view. By reconstructing with 256-pixel patches and leveraging HMFPM's aberration-correction capability, these field-dependent aberrations can be accurately resolved [Fig. 6(b)], enabling high-quality reconstruction across the entire FOV. Representative zoomed-ins from different regions across the FOV are shown in Figs. 6(c1)–6(e1), with the corresponding APIC results using 88 sequential measurements displayed in Figs. 6(c2)–6(e2) as ground truth. For both central and off-axis patches, the HMFPM reconstructions closely match the APIC results, faithfully revealing pathological structures across different regions of the sample.

HMFPM is also capable of addressing scanning-induced aberrations in whole-slide imaging, thereby eliminating the need for laborious refocusing. Figures 6(f)–6(h) show reconstructions from different FOVs acquired at widely separated positions on the slide. Owing to sample unevenness and glass slide tilt, distant FOVs do not lie on the same focal plane, leading to defocus aberrations, as illustrated by the reconstructed aberrations shown in the lower-left insets. HMFPM effectively corrects these scanning-induced aberrations and yields high-quality reconstructions, with representative zoom-in views provided in Figs. 6(f1)–6(h1). In addition, by combining a low-magnification objective with dark-field spectrum extension, HMFPM provides large FOV coverage and thereby reduces stitching artifacts in whole-slide scanning. These advantages, together with its fast acquisition speed, demonstrate the strong potential of HMFPM for whole-slide phase imaging.

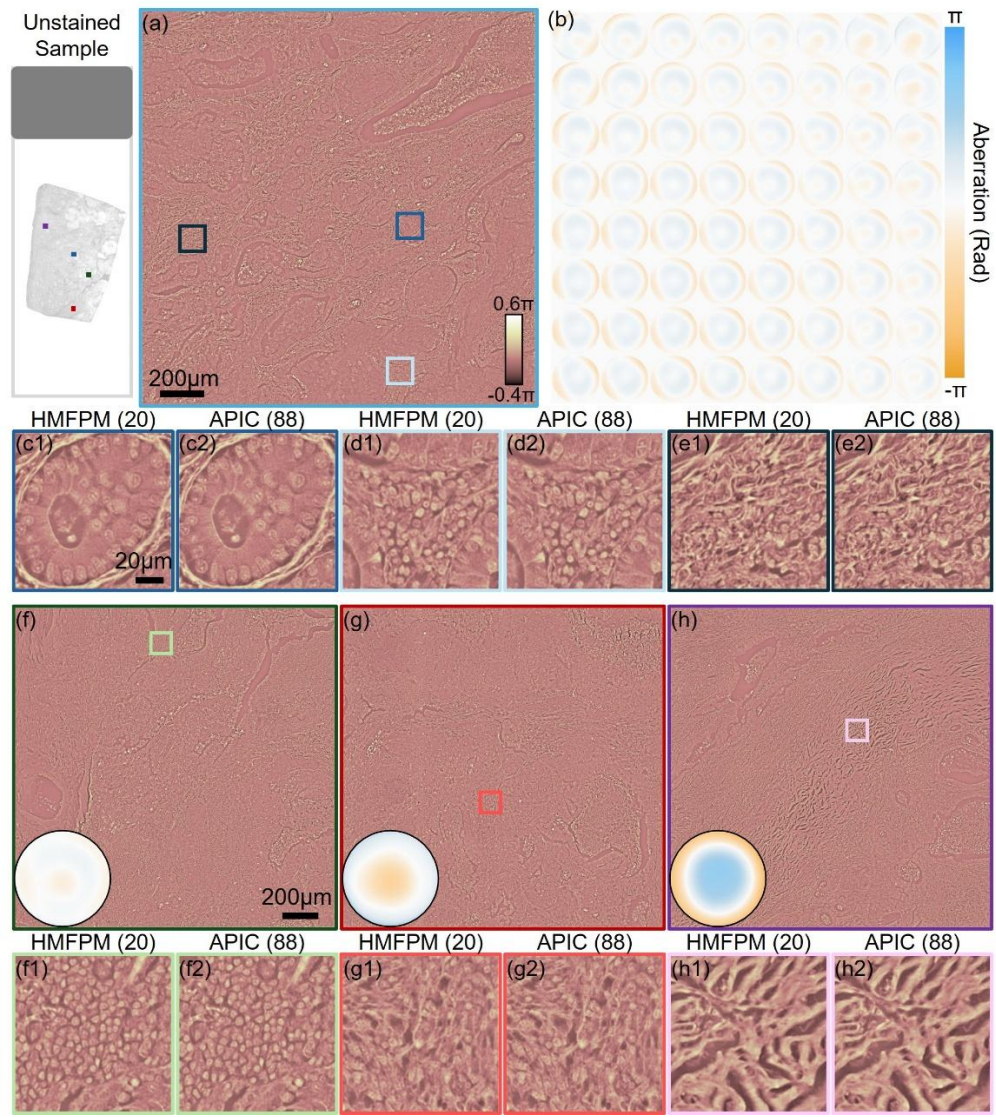

Fig. 6. (a) High-quality phase reconstruction of a representative FOV obtained with HMFPM. (b) Corresponding field-dependent aberrations across the large FOV. (c1–e1) Zoomed-in views of selected regions from (a). (c2–e2) Zoomed-in views of the corresponding regions reconstructed by APIC. (f–h) Phase reconstructions of FOVs at different positions on the whole slide using HMFPM, with reconstructed aberrations shown as insets. (g1–h1) Zoomed-in views of HMFPM results from the cropped regions in (g)–(h). (g2–h2) Zoomed-in views of the corresponding APIC reconstructions.

## 4. Conclusions and Future Work

In this work, we introduced a novel complex-field imaging framework, HMFPM, that improves the efficiency of multiplexed FPM and achieves robust aberration correction. By applying analytic bright-field reconstruction and pupil recovery, followed by customized multiplexed dark-field reconstruction, HMFPM enables high-resolution (1.08 μm, 4-fold improvement over the system's coherent diffraction limit) and wide-field (1.77×1.77 mm$^2$ FOV) imaging with only 20 measurements. Both simulations and experiments verified its robustness under diverse aberrations, including up to 84 μm defocus in experiments, and demonstrated its effectiveness in pathological phase imaging for correcting field-dependent and scanning-induced aberrations.

Despite these advancements, several aspects of the current HMFPM system remain to be improved. First, although multiplexing reduces acquisition time, the overall speed is still limited for applications requiring high temporal resolution. With current setup, HMFPM was able to capture a FOV of 1.77 mm × 1.77 mm within 5 s, which can be further reduced using hardware triggering and brighter LEDs or lasers [43]. With potential hardware improvements, we anticipate that the image acquisition of a FOV of 1.77 mm × 1.77 mm at 1.08 μm resolution can be achieved within 20 ms, equivalent to 50 frame per second video. Furthermore, the acquisition speed can also be improved with better illumination design. The current use of eight NA-matching measurements may be excessive and could potentially be reduced to 4–6 with optimized illumination angles, while multiplexing strategies could also be explored for NA-matching measurements. For dark-field imaging, more efficient multiplexing strategies could further enhance acquisition efficiency.

Second, although HMFPM demonstrates strong aberration-correction capability, its performance degrades under very large defocus (e.g., >100 μm), where accurate aberration extraction becomes challenging. While increasing the number of NA-matching measurements could mitigate this, it would also reduce acquisition efficiency. Alternatively, a recently proposed digital refocusing framework [54] has shown that the digitally summed Fourier spectrum of two images acquired under two-angle illumination produces interference-like fringes at large defocus distances. These fringes can be used to estimate defocus through a physics-based relation, providing a useful defocus aberration prior for subsequent reconstruction. Integrating such a strategy into HMFPM could significantly enhance its aberration-extraction capacity and extend its robustness under stronger defocus conditions.

Building on the foundation of HMFPM, these potential improvements point toward future developments that can further enhance both efficiency and robustness. Ultimately, HMFPM aims to provide a practical, high-throughput, high-resolution, and aberration-free solution for biological and biomedical imaging, with potential applications ranging from large-scale tissue pathology to dynamic cell biology.

**Funding.** Heritage Research Institute for the Advancement of Medicine and Science at Caltech (Grant No. HMRI-15-09-01), Rothenberg Innovation Initiative (RI2) in conjunction with the Hagopian Innovation Prize (Grant No. 25570017)

**Acknowledgments.** We would like to thank Prof. Mark Watson and Prof. Richard Cote at School of Medicine, Washington University of St. Louis for providing non-small-cell lung cancer samples. We thank Zhaoxing Gu at Department of Mathematics, Caltech for discussing algorithm design.

**Disclosures.** The authors declare no conflicts of interest.

**Data availability.** The code and example data will be available on GitHub (https://github.com/Magishe/HMFPM.git) and OSF (https://osf.io/68tq2/) upon the acceptance of the manuscript.